# A Hybrid Forecast of Exchange Rate based on ARFIMA, Discrete Grey-Markov, and Fractal Kalman Model

Gol Kim [a], Ri Suk Yun [b]

([a] Center of Natural Science, University of Sciences, Pyongyang, DPR Korea,
E-mail:golkim124@yahoo.com,

[b] Foreign Economic General Bureau, Pyongyang, DPR Korea)

**Abstract.** We propose a hybrid forecast based on extended discrete grey Markov and variable dimension Kalman model and show that our hybrid model can improve much more the performance of forecast than traditional grey Markov and Kalman models. Our simulation results are given to demonstrate that our hybrid forecast method combined with degree of grey incidence are better than grey Markov and ARFIMA model or Kalman methods.

## 1. Introduction

Forecast of exchange rates has been regarded as one of the most challenging applications of modern time series forecasting. Thus, numerous models have been depicted to provide the investors with more precise predictions.

Hybrid forecast is a well-established and well-tested approach for improving the forecasting accuracy. Therefore, the importance of hybrid forecast methods has steadily increased and it acts still on time series forecasting.

Predictability of stock in Shenzhen stock exchange has been researched by Fan Long-zhen (2001). Hybrid Model using SVM, ARMA and Markov model has been researched by Shreya Dubey, Nishchol Mishra(2011) and Ping-Feng Pai, et al(2006) have applied hybrid SVM-regression model for exchange rate prediction. Hybrid prediction model of CNY/USD exchange rate based on Wavelet and S VR has been performed by Fan-Yong Liu(2010).

Grey system theory has been developed for almost 30 years and has obtained many great successes in practical real-life applications. The various novel grey models have been presented ([2-4,12]). Recently, in order to enhance the forecasting performance of the grey model, the hybrid model combined by the grey model and other models have been researched. Grey LS-SVM forecasting model has been proposed by Deqiang Zhou (2011) and the hybrid model using grey-Markov model improved by nonlinear regression has been considered by

Bo Zhang, Junhai Ma (2011). Hybrid method based on Grey neural network and Markov chain( Jun Zhi1, Jianyong Liu(2010)), A Hybrid Grey Markov Prediction Model( Shuyan Chen, et al.(2007)),hybrid model combined GM(1,1) and ARIMA( Guo-Dong Li,et al.(2008)) have been researched.

In order to enhance the forecasting performance of the time series models, the hybrid model combined by the time series models and other models have been advanced many researches. Generalized ARFIMA process with Markov-switching fractional differencing parameter has been researched by Wen-Jen Tsay, Wolfgang Karl Hardle(2009).

The research for improvement in forecasting accuracy using the hybrid model of ARFIMA and feed forward Neural Network has been performed by Cagdas Hakan Aladag, et al.(2012). Forecast of US Dollar/UK Pound exchange rate using ARMA and ARFIMA has been researched by Olanrewaju. I. et al. (2009). Hybrid model to forecast stock close price using proposed ANN Model of Functional Link fuzzy logic neural Model( Kumaran Kumar. J, Kailas(2012)), hybrid model using ARIMA and regression (Madhur Srivastava, et al.(2010))

On the other hand, since Kalman filter model have been proposed in the prediction fields (Kalman R. E (1960)), now, this prediction model have being used in the financial prediction fields (Fore example, see H. Haleh, B. Akbari et al. (2011))

This paper deals with the development of an improved prediction model using ARFIMA model ,grey method, Markov chain models and the variable fractal dimension kalman and its application to financial time series forecasting. Our method includes *non homogenous discrete grey model, fuzzy weight Markov model, the variable fractal dimension kalman model, hybrid method based on least square weight , hybrid method based on the effective degree weight , hybrid method based on the optimal grey relation degree weight.*

We demonstrate the advantage of our method compared with the traditional grey-Markove and Kalman model through forecasting simulations.

## 2. ARFIMA Model

ARFIMA models are used to model long range dependent time series. ARFIMA models were introduced by [10]. ARFIMA $(p,q,d)$ model with mean $\mu$ can be given by

$$\phi(B)(1-B)^d(X_t - \mu) = \theta(B)e_t, \; -1/2 < d < 1/2$$

where $B$ is the back-shift operator such that $BX_t X_{t-1}$ and $e_t$ is a white noise process with $E(e_t) = 0$ and variance $\sigma_e^2$. The polynomials $\phi(B) = (1 - \phi_1 B - \cdots - \phi_p B^p)$ and

$\theta(B) = (1 - \theta_1 B - \cdots - \theta_q B^q)$ have orders *p* and *q* respectively with all their roots outside the unit circle.

Researches about the parameter estimation of ARFIMA models still continue. The maximum likelihood (ML) methods for ARFIMA are proposed in many literature(fore example [18]; exact ML method (EML) by [22]).

## 3. Discrete grey- Fuzzy and weight Markov model
### 3.1. Non homogenous discrete grey model

Let $x^{(0)} = (x^{(0)}(1), x^{(0)}(2), \cdots x^{(0)}(i), \cdots, x^{(0)}(N))$ be an original sequence and $x^{(1)} = (x^{(1)}(1), x^{(1)}(2), \cdots x^{(1)}(i), \cdots, x^{(1)}(N))$ be accumulated generating sequence by AGO. Where $x^{(1)}(k)$ is given such as:

$$x^{(1)}(k) = \sum_{i=1}^{k} x^{(0)}(i)$$

Then, we define non-homogenous discrete grey model (DGM) such as [[2]];

$$\begin{cases} \hat{x}^{(1)}(k+1) = \beta_1 \hat{x}^{(1)}(k) + \beta_2 \hat{x}^{(1)}(0) + \beta_3 k + \beta_4 \\ \hat{x}^{(1)}(1) = \xi \end{cases}$$

where $\hat{x}^{(1)}(k)$ is fitting value of original sequence and $\beta_1, \beta_2, \beta_3, \beta_4$ are parameters of system.

By least square method, the parameters of system $\beta = [\beta_1, \beta_2, \beta_3, \beta_4]$ are given by

$$\beta = [\beta_1, \beta_2, \beta_3, \beta_4] = (B^T B)^{-1} B^T Y$$

Here, $B = \begin{bmatrix} x^{(1)}(1) & x^{(0)}(1) & 1 & 1 \\ x^{(1)}(2) & x^{(0)}(2) & 2 & 1 \\ \vdots & \vdots & \vdots & \vdots \\ x^{(1)}(n-1) & x^{(0)}(n-1) & n-1 & 1 \end{bmatrix} \quad Y = \begin{bmatrix} x^{(1)}(2) \\ x^{(1)}(3) \\ \vdots \\ x^{(1)}(n) \end{bmatrix}$

The algorithm of non-homogenous discrete grey model (DGM) is given such as;

Step 1; Find the parameters of system $\beta = [\beta_1, \beta_2, \beta_3, \beta_4]$.

Step 2; Put $\hat{x}^{(1)}(1) = \xi$ and find the simulation values $\hat{x}^{(0)}(k), k = 2, 3, \cdots, n$. $\hat{x}^{(0)}(k)$ is the function of $\xi$.

Step 3; Calculate $Q = \sum_{k=1}^{n} [\hat{x}^{(0)}(k) - x^{(0)}(k)]^2$. $Q$ is the function of $\xi$.

Step 4; Put $\frac{dQ}{d\xi} = 0$ and calculate the value $\xi$ which $Q$ have to take minimum.

Step 5; Calculate the value $\hat{x}^{(1)}(k)$ corresponding to $\xi$ obtained from step 4.

Step 6; calculate the forecast value $\hat{x}^{(0)}(k), k = 2, 3, \cdots, n$ by inverse accumulated generating operator (IAGO). That is, $\hat{x}^{(0)}(k) = \hat{x}^{(1)}(k+1) - \hat{x}^{(1)}(k)$

### 3.2. Fuzzy weight Markov model

We assume that $\hat{X}_t$ is the fitting curve which is obtained through forecasting for the time series $Y_t$ by non homogenous discrete grey model.

By considering the actual meaning of the original sequence, we have generated a new time series $Z_t = (Y_t - \hat{X}_t)/Y_{t-1}$ $(t = 2,3,\cdots,N)$ which the fitting curve $\hat{X}_t$ takes as reference

Here $Y_{t-1}$ is 1-period time lag value of $Y_t$

In accordance with the distribution of the random sequence $Z_t$, it is divided by $k$ pieces of state and is taken its partition value. That is,

$$e_1 = [m_0, m_1], e_2 = [m_1, m_2],,\cdots,e_{k-1} = [m_{k-2}, m_{k-1}], e_k = [m_{k-1}, m_k]$$

Here $m_i \geq m_{i-1}$.

A one-step transition probability $P$ is associated with each possible transition from stat $e_i$ to stat $e_j$, and $P$ can be estimated using $P_{ij} = M_{ij}/M_i$ $(i,j = 1,2,\cdots,m)$. $M_i$ means the number of divided pieces whose residuals are stat $e_i$, and $M_{ij}$ is number of transition from $e_i$ to $e_j$ that have occurred by passed one step. These $P_{ij}$ values can be presented as a Transition matrix $R$.

Then, we accept the statistical quantity such as;

$$P_{0j} = \sum_{i=1}^{k} M_{ij} / \sum_{i=1}^{k} M_i, \qquad \chi^2 = 2\sum_{i=1}^{k}\sum_{j=1}^{k} M_{ij} \left|\log \frac{P_{ij}}{P_{0j}}\right|$$

Then, when $N$ is comparatively large, the statistical quantity $\chi^2$ is according to $\chi^2$-distribution with the degree of freedom $(k-1)^2$.

When giving the confidence degree $\alpha$, if $\chi^2 > \chi_\alpha^2(m-1)^2$, then we confirm that the random sequence $Z_t$ have Markov's property, unless, the sequence haven't Markov's property.

Suppose $U$ is the range which random variable of Markov chain takes the value.

We construct the fuzzy state set $S_1, S_2,\cdots,S_l$. If for arbitrary $u \in U$ the condition

$$\sum_{m=1}^{l} \mu_{S_m}(u) = 1$$

is satisfied, and then $\mu_{S_m}(u)$ is called the membership degree of the fuzzy state $S_M$ for numerical value $U$.

Suppose that $\mu_{S_i}(Z_t) \cdot \mu_{S_j}(Z_{t+1})$ is the fuzzy state transition coefficient from the state $S_i$ to $S_j$ when time is turned from $t$ to $t+1$.

Then,

$$a_{ij} = \sum_{t=1}^{N-1} \mu_{S_i}(Z_t) \cdot \mu_{S_j}(Z_{t+1})$$

is called fuzzy transition frequency number from the state $S_i$ to the state $S_j$.

When the state $Z_t$ belongs to $S_i$ with degree of member $\mu_{S_i}(Z_t)$ and belongs to $S_j$ with membership degree $\mu_{S_j}(Z_t)$, the transition order is only expressed by the product between membership degrees $\mu_{S_i}(Z_t) \cdot \mu_{S_j}(Z_{t+1})$.

Owing to the fuzzy transition probability from the state $S_i$ to the state $S_j$ is denoted by

$$P_{ij} = \frac{a_{ij}}{\sum_{j=1}^{} a_{ij}}, ((i,j=1,2,\cdots,M))$$

, the time series time series which we are going to establish is given such as;

$$\hat{Y}_t = \hat{X}_t + \sum_{i=1}^{k} \mu_{S_i}(Z_{t-1}) \sum_{j=1}^{k} \frac{1}{2}(m_{i-1} + m_i) P_{ij} Y_{t-1}, \quad t = 1,2,\cdots,N$$

Here, $\hat{X}_t$ is the predicted value obtained by using non homogenous discrete grey model.
The forecasting model obtained by above methods is called the non homogenous discrete grey -fuzzy weight Markov model. We denote this model by DGM-FMarkov.

## 4. Fractal Kalman Model

The sequence $\{\lg(\frac{N_i}{N_{i+1}})\}$ $(i = 1,2,3,\cdots)$ is called the observation series and it denote by $\{x_i\}$ $(i = 2,3,4,\cdots)$. Here $N_i$ is actual exchange rate value theoretically.

By statistical noise, the exchange rate value obtained practically is $\hat{N}_i$.

The sequence $\{\lg(\frac{r_i}{r_{i+1}})\}$ $(i = 1,2,3,\cdots)$ is called the time series and it denote by $\{t_i\}$ $(i = 2,3,4,\cdots)$. Here $t_i = \lg(i + \frac{1}{i})$ because $r_i = i$.

The sequence $\{\frac{x_i}{t_i}\}$ $(i = 1,2,3,\cdots)$ is called the fractal degree series and it denote by $\{d_i\}$ $(i = 2,3,4,\cdots)$.

From here, we obtain such as;

$$t_{n+1} = \log\frac{r_{n+1}}{r_n} = \lg\frac{r_n}{r_{n-1}}\frac{r_{n-1}}{r_n}\frac{r_{n+1}}{r_n} = t_n + \lg\frac{n^2-1}{n^2}$$

Therefore, we obtain the expression such as;

$$x_{n+1} = d_{n+1} t_{n+1} = (d_n + \Delta d_n) \cdot (t_n + \lg(n^2 - 1/n^2))$$
$$= d_{n+1} t_{n+1} + \lg(n^2 - 1/n^2)) \cdot d_n + \lg(n/n-1) + \lg(n^2 - 1/n^2)) \cdot \Delta d_n$$
$$= x_n + \lg(n^2 - 1/n^2)) \cdot d_n + \lg(n/n-1) \cdot \Delta d_n$$

Here $\Delta d_n = d_{n+1} - d_n$ and it is called by the increment of fractal degree.

We can regard that the increment of fractal degree $\Delta d_n$ is zero-mean white noise.

Therefore, from above expression, we can build the state equation and model observation equation of Kalman filter in exchange rate market.

$$\begin{bmatrix} x_{k+1} \\ d_{k+1} \end{bmatrix} = \begin{bmatrix} 1 & \lg(k^2 - 1/k^2) \\ 0 & 1 \end{bmatrix} \begin{bmatrix} x_k \\ d_k \end{bmatrix} + \begin{bmatrix} \lg(k^2 - 1/k^2) \\ 1 \end{bmatrix} \Delta d_k$$

$$z_k = \begin{bmatrix} 1 & 0 \end{bmatrix} \cdot \begin{bmatrix} x_k \\ d_k \end{bmatrix} + V_k$$

Here, $z_k = \lg(\hat{N}_k / \hat{N}_{k+1})$ and $V(k)$ is observation noise.

It is statistical error of trade member in the exchange rate transaction. It can be seen as zero-mean white noise. Now, we put such as;

$$X(k+1) = \begin{bmatrix} x_{k+1} \\ d_{k+1} \end{bmatrix}, \quad \Phi(k+1,k) = \begin{bmatrix} 1 & \lg(k^2 - 1/k^2) \\ 0 & 1 \end{bmatrix}, \quad W(k) = \Delta d_k,$$

$$\Gamma(k+1,k) = \begin{bmatrix} \lg(k^2 - 1/k^2) \\ 1 \end{bmatrix}, \quad C_k = \begin{bmatrix} 1 & 0 \end{bmatrix}, \quad Z_k = z_k$$

Therefore, we can obtain matrix form equation such as;

$$\begin{cases} X(k+1) = \Phi(k+1,k) X(k) + \Gamma(k+1,k) W(k) \\ Z(k) = C(k) X(k) + V(k) \end{cases}$$

The fundamental formula of Kalman filter recurrence algorithm is given by such as;

$$\begin{cases} \hat{X}(k+1 | k+1) = \hat{X}(k+1 | k) + K(k+1) \cdot \varepsilon(k+1) \\ \hat{X}(k+1 | k) = \Phi(k+1,k) \cdot \hat{X}(k | k) \\ \varepsilon(k+1) = Z(k+1) - C(k+1) \hat{X}(k+1 | k) \\ K(k+1) = P(k+1 | k) C^T(k+1) [C(k+1) P(k+1 | k) C^T(k+1) + R(k+1)]^{-1} \\ P(k+1 | k) = \Phi(k+1 | k) P(k | k) \Phi^T(k+1,k) + \Gamma(k+1,k) Q(k) \Gamma^T(k+1,k) \\ P(k | k) = [I_n - K(k) C(k)] P(k | k-1) \end{cases}$$

The initial conditions of Kalman filter are given by such as;

$$\Phi(2,1) = \begin{bmatrix} 1 & \lg(1/4); & 0 & 1 \end{bmatrix}, \quad \Gamma(2,1) = \begin{bmatrix} \lg(1/2) & 1 \end{bmatrix}^T, \quad Q_k \equiv 0.0001,$$

$R_k \equiv 1.0$ , $\hat{X}(1/1) = [x_2 \quad d_2]^T = [\lg(N_1/N_2) \quad \lg(N_1/N_2)/\lg 2]$ , $P(1/1) = 0.0001 \times I_2$ .

Here $I_2$ is two-dimension unit matrix.

From filter recurrence algorithm, we can obtain one-step predicted value $\hat{X}(n+1/n)$.

The component of it is given such as; $\hat{x}(n+1/n) = \lg(N_n/\hat{N}_{n+1})$. Here $n$ is the given sample number.

Therefore, the forecast value of exchange rate in time $r_{n+1}$ is given such as;

$$\hat{N}_{n+1} = N_n / 10^{\hat{x}(n+1/n)}$$

We denote variable fractal dimension Kalman model by F-Kalman.

## 5. The hybrid forecasting model

Determination method of combinational weights in construction of hybrid forecast model is very important problem. The past, in determination methods of combinational weights, there are such as simple average(AVG), Variance based(VAR), Inverse of the mean square error (INV-MSE), Rank based weighting (RANK) ,Least squares estimation ,Shrinkage method (SHRINK) ,Geometric mean ,Harmonic mean, a method based on testing the performance difference, Hierarchical forecast combination (HIER) ( Robert R.et al. (2011) ) ect.

In this paper we have considered the determination method of combinational weights by based on least square method, based on the effective degree method, based on optimal grey relation degree method

### 5.1. The determination of combinational weight based on least square method

Let $y = \{y(1), y(2), \cdots, y(N)\}$ be original series and $\hat{y}_j(t)$ $(j=1,2,\cdots;m)$ be predicted value of the $j$'th method in $t$ time. And let $w_j$ $(j=1,2,\cdots;m)$ be its weights, which satisfies as following condition; $\sum_{j=1}^{m} w_j = 1$ , $w_j \geq 0$ $(j=1,2,\cdots;m)$

Now, we construct the hybrid forecast model such as;

$$\hat{y}(t) = \sum_{j=1}^{m} w_j \hat{y}_j(t)$$

We introduce the notation such as

$$H = diag[h_{ij}], \quad W = [w_1, w_2, \cdots, w_m]^T, \quad E = [1,1,\cdots,1]^T$$

The weight based on least square method has been obtained such as

$$\min J = W^T H W$$

$$s.t. \begin{cases} E^T W = 1 \\ W \geq 0 \end{cases}$$

Solving these equations Lagrange multiplier method, we obtain such as;

$$w_j = \frac{1}{h_{jj} \sum_{i=1}^{m} \frac{1}{h_{ii}}}$$

Here $h_{jj} = \sum_{t=1}^{n} \varepsilon_{jt}^2$ and $\varepsilon_{jt} = \hat{y}_j(t) - y_j(t)$, $(j = 1, 2, \cdots, m; t = 1, 2, \cdots, N)$.

### 5.2. The determination of combinational weight based on the effective degree method

To explain a discussion briefly, let us $\hat{y}_i(t)$ $(t = 1, 2, \cdots N, (i = 1, 2, \cdots m)$ are the predicted value obtained by $i$'th kind of forecast method.

Now, if we introduce

$$A(t) = 1 - \left| \frac{y(t) - \hat{y}(t)}{y(t)} \right| = 1 - \left| \frac{y(t) - \sum_{i=1}^{m} k_i \hat{y}_i(t)}{y(t)} \right|$$

then $A(t)$ constitutes the accuracy of the hybrid forecast

Let us define the average value $E$ and average square deviation $\sigma$ of $A(t)$ such as respectively

$$E = \frac{1}{N} \sum_{t=1}^{N} A(t), \quad \sigma = \frac{1}{N} (\sum_{t=1}^{N} (A(t) - E)^2)^{\frac{1}{2}}$$

Then, the effective degree of the comprehensive forecast method has been defined such as

$$S = E(1 - \sigma)$$

It show that the greater $S$ is, the more higher the accuracy of the forecast model, the more stable and the more and more effective the model

Therefore, we have to find out the weights $k_1$ and $k_2$ by optimizing of $S$. That is

$$\min S = E(1 - \sigma)$$

### 5.3. The determination of combinational weight based on optimal grey relation degree method

Let $\{y(k) | k = 1, 2, \cdots, N\}$ be original value of time series, $\hat{y}_j(k)$ $(i = 1, 2, \cdots, m, k = 1, 2, \cdots, N)$ be predicted value of the $j$'th forecast method in $k$ time $(j = 1, 2, \cdots, m, t = 1, 2, \cdots, N)$.

We put

$$\gamma_{0j} = \frac{1}{N} \sum_{t=1}^{N} \frac{\min_{1 \leq j \leq m} \min_{1 \leq t \leq N} |e_j(t)| + \rho \max_{1 \leq j \leq m} \max_{1 \leq t \leq N} |e_j(t)|}{|e_j(t)| + \rho \max_{1 \leq j \leq m} \max_{1 \leq t \leq N} |e_j(t)|}$$

Then, $\gamma_{0j}$ is called the grey relation degree of the predicted value sequence

Here $\rho \in (0,1)$ is called the identification coefficient. Generally we take $\rho = 0.5$.

$e_j(t) = y(t) - \hat{y}_j(t)$ is the predicted error of $k$ time for $i$'th forecast method.

Let $\hat{y}(t) = \sum_{j=1}^{m} w_j \hat{y}_j(t)$ be the predicted value of $y(t)$ by the hybrid forecast method.

Here $w_1, w_2, \cdots, w_m$ is the weight coefficient of $m$ kind of forecast method, which satisfies $\sum_{j=1}^{m} w_j = 1, w_j \geq 0 \ (j=1,2,\cdots,m)$.

The combinational weight based on the optimal grey relation degree is determined such as;

$$\max \gamma(W) = \frac{1}{N} \sum_{t=1}^{N} \frac{\min_{1\leq j\leq m} \min_{1\leq t\leq N} |e_j(t)| + \rho \max_{1\leq j\leq m} \max_{1\leq t\leq N} |e_j(t)|}{\left|\sum_{j=1}^{m} w_j e_j(t)\right| + \rho \max_{1\leq j\leq m} \max_{1\leq t\leq N} |e_j(t)|}$$

$$s.t. \begin{cases} \sum_{j=1}^{m} w_j = 1 \\ w_j \geq 0, \ j=1,2,\cdots,m \end{cases}$$

By solving this optimal problem, we can determine the combinational weight $W = \{w_1, w_2, \cdots, w_m\}^T$ based on the optimal grey relation degree.

### 5.4. Hybrid forecast

Let $\hat{y}_1(t), \hat{y}_2(t), \cdots, \hat{y}_m(t)$ be predicted values of $m$ sort of another model in time $t$

Let $\mu_j \ (j=1,2,\cdots,m)$ be the weight of $j$'th model obtained by our methods of determination of combinational weight. Then, the hybrid forecast model added weight is given by

$$\hat{y}(t) = \sum_{j=1}^{m} \mu_j \hat{y}_j(t), \ \sum_{j=1}^{m} \mu_j = 1$$

First, we have combined ARFIMA, DGM-FMarkov and F-Kalman by weight based on least square, which is called by method 1.

Then, we have combined ARFIMA, DGM-FMarkov and F-Kalman by weight based on the effective degree, which is called by method 2.

Third, we have combined ARFIMA, DGM-FMarkov and F-Kalman by weight based on the optimal grey relation degree, which is called by method 3.

## 6. Forecast simulation experiments

To compare the forecast performances of the models the monthly seasonally adjusted US dollar-UK pound exchange rate data covering from January 1971 to December 2008 (T = 456) is used. These data were obtained from IMF International financial Statistics and which we believe is large enough to model the fractional integrated ARMA model. The forecast simulation has been performed by MATLAB 7.8 (2009a)

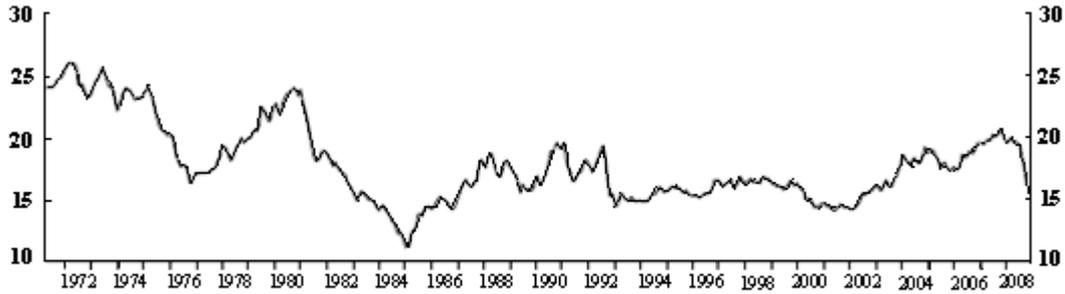

Figure 1: Time Plot of UK Pound to US Dollar Exchange Rate Jan.1991-Dec.2008

By using the AIC and the probability of significance of the parameter estimate are used to select the best fitting model. The best model is the zero mean $ARFIMA(3,d,0)$ model with parameters $d = 0.4957$, $\phi_1 = 0.9456$, $\phi_2 = -0.2665$, $\phi_1 = 0.1953$,

$$(1 - 0.9456B + 0.2665B^2 - 0.1953B^3)(1-B)^{0.4957} X_t = \varepsilon_t$$

The optimal forecasts values are then evaluated using the mean absolute forecast error (MAFE) defined as,

$$MAFE = \frac{1}{h+1} \sum_{t=s}^{h+s} \left| \hat{X}_t - X_t \right|$$

the root mean square forecast error (RMSFE) defined as:

$$RMSFE = \sqrt{\frac{1}{h+1} \sum_{t=s}^{h+s} (\hat{X}_t - X_t)^2}$$

and the mean absolute percentage forecast error (MAPFE) is also given as,

$$MAPFE = \frac{100}{h+1} \sum_{t=s}^{h+s} \left| \frac{\hat{X}_t - X_t}{\hat{X}_t} \right|$$

where $t = s, 1+s, \cdots, h+s$, $h$ is denote the forecast prediction length and the actual and predicted value for corresponding $t$ values are denoted by $\hat{X}_t$ and $X_t$ respectively.

The directional accuracy(DA) defined as

$$DA = \frac{100}{h+1} \sum_{t=s}^{h+s} d_t, \quad d_t = \begin{cases} 1, & (X(t) - X(t-1))(\hat{X}(t) - X(t-1)) \geq 0 \\ 0, & otherwise \end{cases}$$

The smaller the values of RMSFE and MAPFE, the better the forecasting performance of the model. The lager the value of DA, the better the forecasting performance of model.

The results of prediction are shown in the following table . In our simulation experiment, $h = 12$

Table . Prediction Performance

| Method | Evaluate criteria | | | |
|---|---|---|---|---|
| | MAPFE | DA | MAFE | RMSFE |
| Method of [10] | 6.0833 | 63.6 | 0.10527 | 0.1930 |
| Method 1 | 4.0627 | 72.8 | 0.09573 | 0.1573 |
| Method 2 | 3.8763 | 72..8 | 0.08692 | 0.1265 |
| Method 3 | 2.8697 | 81.8 | 0.07235 | 0.0976 |

Here Method of [10] is ARFIMA model, which has been researched in paper [10].

From this table, we see that the methods 3 are better than other methods in criterion RMSE. The method 3 is best from the view point of DA, and the method 3 is best from the view point of MAPFE, MAFE and RMSFE. In particular, the method 3, which is by weight based on the optimal grey relation degree, has the best performance. This shows that hybrid forecast is better than single one.

## 7. Conclusion

We proposed hybrid forecast method based on ARFIMA, non homogenous discrete grey-Markov and variable fractal dimension Kalman and showed that our hybrid model can improve much the forecast performance than traditional grey-Markov and Kalman models. Our simulation demonstrated that hybrid model combined ARFIMA, DGM-FMarkov and F-Kalman by weight based on varied method are much better than single method. Especially, the method ARFIMA, DGM-FMarkov and F-Kalman combined by weight based on the optimal grey relation degree showed the best performance.

School of Management, Fudan University